\documentclass[a4paper,aps,showpacs,nofootinbib,amsmath]{revtex4}
\usepackage{epsfig}
\usepackage{color}
\usepackage{amssymb}
\usepackage{amsfonts}
\usepackage{graphicx}
\usepackage{keyval,graphicx}
\usepackage{textcomp,wasysym}
\usepackage{slashed}

\begin{document}

\bibliographystyle{unsrt}

\title{Understanding the newly observed heavy pentaquark candidates}

\author{Xiao-Hai Liu$^1$\footnote{liuxh@th.phys.titech.ac.jp}, Qian Wang$^2$\footnote{q.wang@fz-juelich.de},
and Qiang Zhao$^3$\footnote{zhaoq@ihep.ac.cn}}

\affiliation{$^1$ Department of Physics, H-27, Tokyo Institute of Technology, Meguro, Tokyo 152-8551, Japan}

\affiliation{$^2$   Institut f\"{u}r Kernphysik, Institute for Advanced Simulation, and J\"ulich Center for Hadron
Physics, D-52425 J\"{u}lich, Germany }

\affiliation{$^3$ Institute of High Energy Physics and Theoretical Physics Center for Science Facilities,
        Chinese Academy of Sciences, Beijing 100049, China}

\date{\today}

\begin{abstract}

We find that several thresholds can contribute to the enhancements of the newly observed heavy pentaquark candidates $P_c^+(4380)$ and $P_c^+(4450)$ via the anomalous triangle singularity (ATS) transitions in the specific kinematics of $\Lambda_b\to J/\psi K^- p$. Apart from the observed two peaks we find that another peaks around 4.5 GeV can also be produced by the ATS. We also show that the $\Sigma_c^{(*)}$ can be produced at leading order in $\Lambda_b$ decay. This process is different from the triangle diagram and its threshold enhancement only appears as CUSP effects if there is no pole structure or the ATS involved. The threshold interaction associated with the presence of the ATS turns out to be a general phenomenon and plays a crucial role in the understanding of candidates for exotic states.

\end{abstract}



\maketitle

\section{Introduction}

The two states $P_c^+(4380)$ and $P_c^+(4450)$ observed in the invariant mass spectrum of $J/\psi p$ in $\Lambda_b\to J/\psi K^- p$ by the LHCb Collaboration~\cite{Aaij:2015tga} has immediately attracted a lot of attention from the whole community since they could be the long-searching-for pentaquark states in the heavy flavor sector. Recent theoretical studies can be found in Refs.~\cite{Wu:2010jy,Wu:2010vk,Wu:2012md,Yang:2011wz,xiao-2013}. Their masses are $4380\pm 8 \pm 29$ MeV and $4449.8\pm 1.7\pm 2.5$ MeV, respectively, and their widths are $205\pm 18\pm 86$ MeV and $39\pm 5\pm 19$ MeV, respectively. Their preferred $J^P$ are either $3/2^-$ and $5/2^+$, or $3/2^+$ and $5/2^-$, respectively, based on the detailed experimental analysis. The possible existence of multiquark states has always been regarded as a natural consequence of QCD. Although the conventional quark model has made tremendous successes in the description of the hadron spectroscopy, it also raised questions on why and how those multiquark states kept out of our sight for such a long time. The LHCb results and the recent results from Belle~\cite{Choi:2007wga} and BESIII~\cite{Ablikim:2013mio} certainly make a big step forward to our understanding the multiquark system. But at the same time, they also give us chances to ask more questions.

In this Letter we will examine the role of the non-perturbative anomalous triangle singularity (ATS) in the decay of $\Lambda_b\to J/\psi K^- p$. As recently pointed out in Ref.~\cite{Guo:2014iya}, the pronounced narrow threshold states in the elastic channels should indicate non-perturbative rescatterings which will eventually generate pole structures after sum over all the loops to infinity. It was also stressed that if the ATS is present the threshold peak will be enhanced and mix with the dynamic pole structure in the inelastic channel. Therefore, in order to understand the nature of the threshold enhancements a careful analysis of the triangle process should be necessary.

There are several interesting features arising from the decay channel as the data have shown. For instance, there are clear structures for $\Lambda^*$ resonances in the lower end of the $K^- p$ invariant mass spectrum while the higher mass region appear to smooth out. In contrast, the observed  $P_c^+(4380)$ and $P_c^+(4450)$, in particular, the  $P_c^+(4450)$, are clear structures above the phase space. Note that some interferences occur over a rather broad mass range above the  $P_c^+(4450)$ in the $J/\psi p$ invariant mass spectrum (see e.g. Fig.~2 of Ref.~\cite{Aaij:2015tga}). This feature suggests that there are intermediate processes in $\Lambda_b\to J/\psi K^- p$ which will give rise to final state interactions. Our motivation is to investigate whether the ATS will accumulate the events at the two peak positions.

As follows, in the next Section we will first analyze the transition mechanisms for $\Lambda_b\to J/\psi K^- p$ and then investigate the loop diagrams where the ATS can be present. A summary will be given in the last Section.

\section{Production mechanisms for threshold states}

The transition of $\Lambda_b\to J/\psi K^- p$ is dominated by the flavor changing weak decay of $b\to c + \bar c s$ via the $V-A$ current. This leads to an intuitive expectation that the $u$ and $d$ quark in $\Lambda_b$ should be a spectator, hence their isospin quantum number $I=0$ should be conserved.  As a consequence, if the $ud$ pair is to be combined with a $c$ quark to form a charmed baryon, it will favor $\Lambda_c$ baryons instead of $\Sigma_c$ ones.
In Refs.~\cite{Chen:2015loa,Chen:2015moa} the authors propose that $P_c^+(4380)$ and $P_c^+(4450)$ are molecular states of $\Sigma_c(2455) {\bar D}^*(2007)$ and $\Sigma_c(2520){\bar D}^*(2007)$. But it was not discussed why the production of $\Sigma_c^{(*)}$ in the $\Lambda_b$ decay should be significant.  Also, in Ref.~\cite{Karliner:2015ina} the authors refer one of the pentaquark candidates to a molecular state of $\Sigma_c(2455) {\bar D}^*(2007)$ with $J^P=3/2^-$. In Ref.~\cite{Roca:2015dva}  the authors proposed to produce the $P_c^+(4450)$ via  the $J/\psi p$ rescattering to $\Sigma_c^{(*)} {\bar D}^{(*)}$ which, however, should be higher order terms.  In a followed-up paper~\cite{Feijoo:2015cca} the authors investigate other isospin 0 channel recoiled by $J/\psi$.

\begin{figure}
\centering
\includegraphics[width=0.6\textwidth]{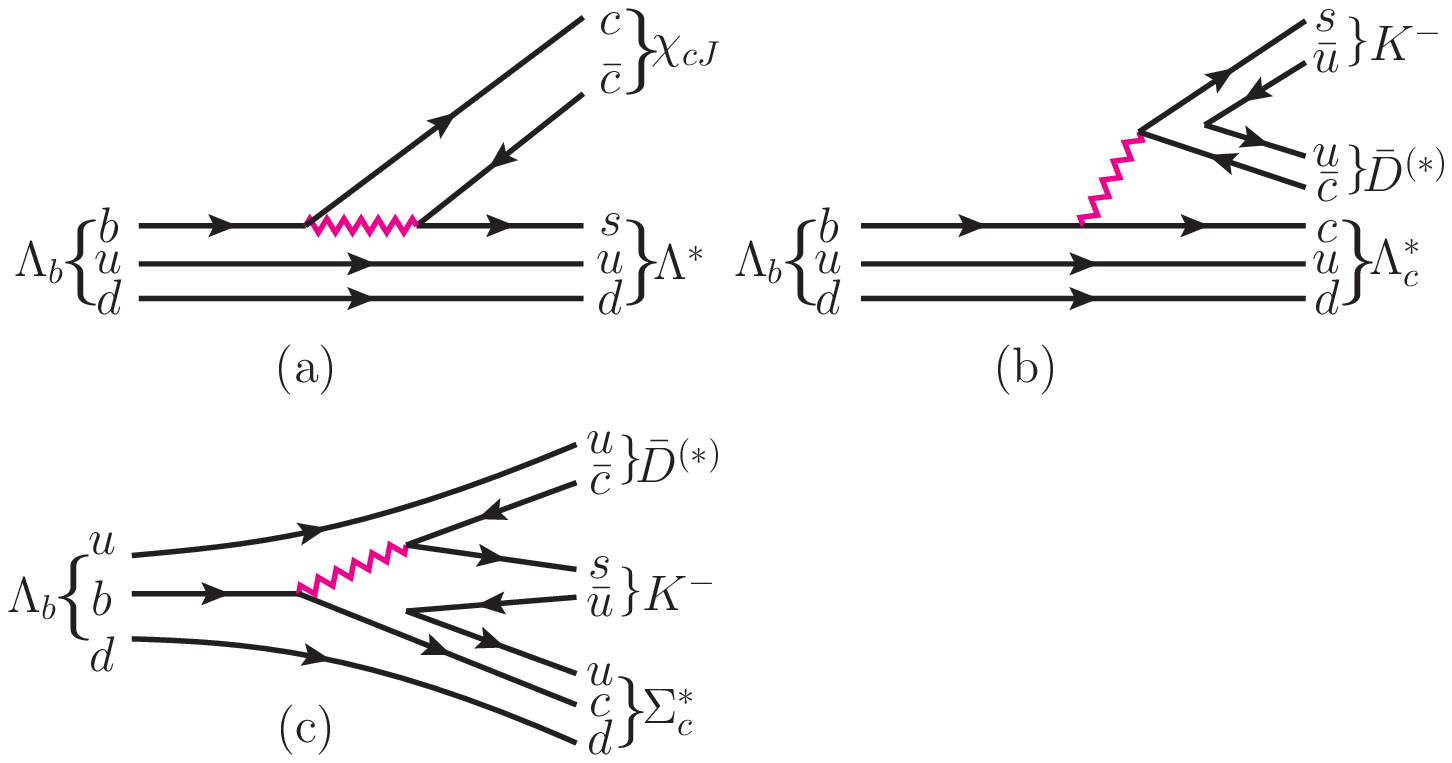}
\caption{Relevant processes for the production of $P_c^+(4380)$ and $P_c^+(4450)$ in $\Lambda_b\to J/\psi K^- p$.  }
\label{fig-1}
\end{figure}

Actually, the production of the $\Sigma_c^{(*)}$ is not necessarily to be suppressed at leading order. Since the $b$ quark decay is a short-distance process, the final $c$, $\bar c$ and $s$ quark will be created locally. They can easily split the $ud$ quark and rearrange them with the created $u\bar u$ to form, e.g. $\Sigma_c^{(*)} {\bar D}^{(*)} K^-$. Then the rescattering of $\Sigma_c^{(*)} {\bar D}^{(*)}$ into $J/\psi p$ can occur. This process is illustrated in Fig.~\ref{fig-1} (c) with (a) and (b) the same mechanism analyzed in Ref.~\cite{Aaij:2015tga} as a contrast.

For Fig.~\ref{fig-1} (b), a decomposition of the isospin gives the relative decay amplitudes
\begin{eqnarray}\label{trans-1}
&&\langle Y_c \bar K \bar D | {\hat H}_w | \Lambda_b\rangle_{(b)}  \nonumber\\
&= & \frac{1}{\sqrt{2}}( \Lambda_c^+ {\bar K}^0 D^- -\Lambda_c^+ K^- {\bar D}^0 ) \ ,
\end{eqnarray}
where $Y_c$ denotes charmed baryon $\Lambda_c^{(*)}$ or $\Sigma_c^{(*)}$. We denote the possible $\Lambda_c^{(*)}$ and ${\bar D}^{(*)}$ by the ground state symbols and simply write the amplitudes by the final state particles. The flavor wavefunction $(ud-du)b/\sqrt{2}$ for $\Lambda_b$ is implied.

Similarly, the decomposition for Fig.~\ref{fig-1} (c) can be written as
\begin{eqnarray}\label{trans-2}
&&\langle Y_c \bar K \bar D | {\hat H}_w | \Lambda_b\rangle_{(c)}  \nonumber\\
&= & \frac{1}{2\sqrt 2} \left[ -  \Sigma_c^{++} K^- D^- +\frac 12 \Sigma_c^+{\bar K}^0 D^- -\frac{1}{2} \Sigma_c^+ K^- {\bar D}^0 +  \Sigma_c^0{\bar K}^0 {\bar D}^0 +\frac 12 \Lambda_c^+ K^- {\bar D}^0 -\frac 12 \Lambda_c^+{\bar K}^0 D^- \right] \ ,
\end{eqnarray}
where $\Sigma_c^{++} K^- D^-$ and $\Sigma_c^+ K^- {\bar D}^0$ would allow the formation of pentaquark states via the $\Sigma_c^{(*)} \bar D^{(*)}$ rescatterings as studied in Refs.~\cite{Chen:2015loa,Chen:2015moa,Karliner:2015ina,Roca:2015dva}. From Eqs.~(\ref{trans-1}) and (\ref{trans-2}) we can see that Fig.~\ref{fig-1} (b) dominantly contribute to the production of $I=1/2$ pentaquark system recoiled by $K^-$.

\begin{figure}
\centering
\includegraphics[width=0.5\textwidth]{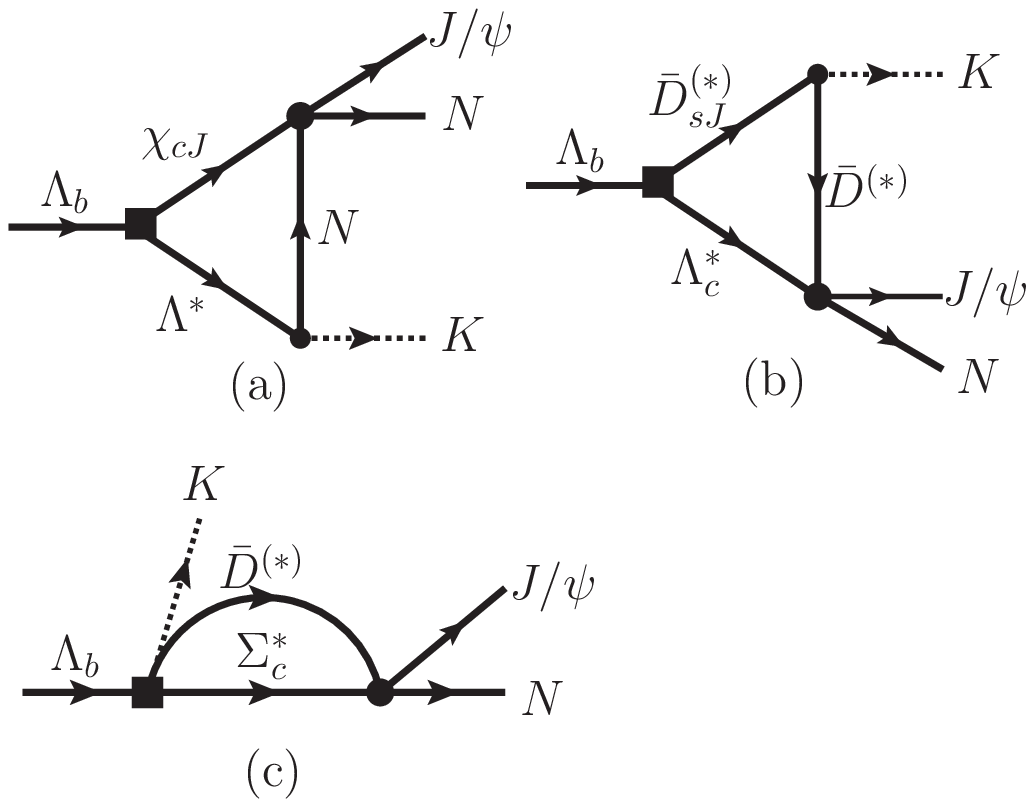}
\caption{The loop diagrams as a consequence of Fig.~\ref{fig-1} where the ATS and kinematic CUSP can be recognized.  }
\label{fig-2}
\end{figure}

If the production of the pentaquark candidates $P_c^+(4380)$ and $P_c^+(4450)$ is indeed via the $\Sigma_c^{(*)}\bar D^{(*)}$ (Table~\ref{tab-3}) or $\Lambda_c^{(*)}\bar D^{(*)}$ (Table~\ref{tab-1} ) interactions, then the relevant thresholds will play non-trivial roles in the coupled-channel decays into e.g. $J/\psi p$. Considering the major rescattering processes arising from Fig.~\ref{fig-1}, we can reexpress the transitions as Fig.~\ref{fig-2} where three type of rescatterings will contribute. Figure~\ref{fig-2} (a) and (b) are triangle diagrams while (c) is two-point loop interactions.  We note that the creation of $\Sigma_c^{(*)}$ via Fig.~\ref{fig-1} (c) should be driven by short-distance interactions.  This process will favor the formation of $P_c^+(4380)$ and $P_c^+(4450)$ if they are states generated by $\Sigma_c^{(*)}\bar D^{(*)}$ interactions.

The interesting property of Fig.~\ref{fig-2} (a) and (b) is that given the masses of the involved states located within certain ranges it will allow the internal states to be on-shell simultaneously. This is different from the kinematic CUSP effects which only recognizes the on-shell condition for two internal particles and contributions from such a branch point is subleading compared to the ATS~\cite{Landau:1959fi,bonnevay:1961aa}. Therefore, we do not expect that such CUSP effects produce narrow and strongly enhanced structures in the invariant mass spectrum.  In contrast, when the ATS condition is satisfied, the singularity behavior of the integral will produce strong enhancements at the singular points of which the effects can be measured in the experiment. In particular, the singular points will mostly locate in the vicinity of the two-body thresholds but not necessarily to be exactly at the thresholds. It should be realized that the positions of the singularity will not change even when higher partial waves contribute at the interaction vertices.\footnote{ The only difference is that the higher partial waves will make the signal from the ATS less pronounced.} The reason is because the singular term will always be kept in the decomposition of the integrand in the Feynman parametrization. In other words, even though the contribution from the singular term relative to other contributions might be small, its enhancement at the singular point may not be negligible\footnote{The detailed discussion about the ATS and their manifestations in physical processes can be found in Ref.~\cite{Liu:2015taa} and there are cases that the ATS involving higher partial wave interactions can still produce significant threshold enhancements~\cite{Wu:2011yx,Wu:2012pg,Wang:2013cya,Wang:2013hga,Liu:2013vfa,Liu:2014spa}.}. Nevertheless, in the case of $\Lambda_b\to J/\psi K^- p$ there are several thresholds close to each other. Even a small singularity enhancement can build up and produce measurable effects.

Since quite a lot of thresholds can appear in the decays of Fig.~\ref{fig-2} and we are still lack of information about the vertex couplings, we only consider low partial waves and thresholds which are close to the masses of interest and we discuss separately the properties of those three types of loops in Fig.~\ref{fig-2}.

Figure~\ref{fig-2} (a) is a consequence of Fig.~\ref{fig-1} (a) where  the rescattering between $\chi_{cJ}$ and an exchanged proton from the decay $\Lambda^*\to p K^-$ is considered.  Note that the mass thresholds for $p + \chi_{cJ}$ ($J=0, \ 1, \ 2$) are close to the peak masses for $P_c^+(4380)$ and $P_c^+(4450)$ as listed in Table~\ref{tab-chic}. Also, the $S$-wave scatterings of $p \chi_{c2} \to J/\psi p$ \footnote{Although the production of $\chi_{c0}$ and $\chi_{c2}$ is suppressed in $B$ decay or any hadron with only one heavy $b$ quark without charm or anti-charm quark as discussed in Ref.~\cite{Guo:2015umn}, we can still estimate how large the next leading order contributions are or what is the behavior after considering it.} can access the quantum numbers of $3/2^+$ and $5/2^+$ for the threshold enhancement. The $\chi_{c1}$ and $p$ scattering can access the quantum numbers of $1/2^+$ and $3/2^+$. The $\chi_{c0} p$ can reach $1/2^-$ and $3/2^-$ via a $P$ wave interaction. It is interesting to notice that the significant enhancement to the $\chi_{cJ} p$ (with $J=0, \ 1, \ 2)$ via the ATS would prefer that the mass of $\Lambda^*$ within the mass regions $1.92 \sim 2.20 \ \mathrm{GeV}$, $1.89  \sim 2.11 \ \mathrm{GeV} $, $1.83 \sim 2.06 \ \mathrm{GeV}$, respectively. From Fig.~2 (a) of Ref.~\cite{Aaij:2015tga}, it shows that the cross section for $K^-p$ is smooth but non-zero. Note that as long as the kinematics approaching the ATS condition, all the cross sections will contribute to the threshold singularity.

\begin{figure}
\centering
\includegraphics[width=0.5\textwidth]{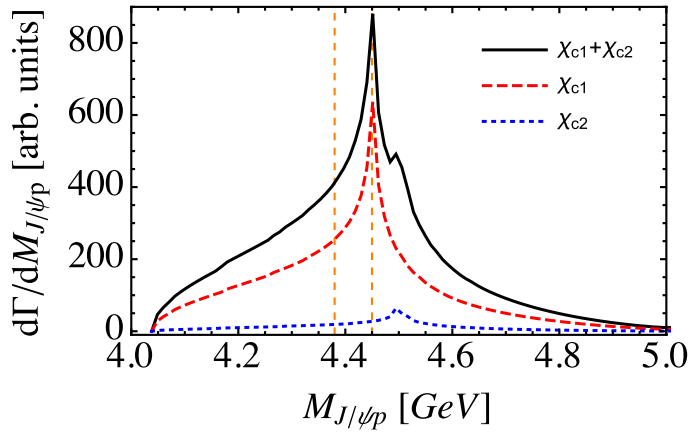}
\caption{The invariant mass distribution of $J/\psi p$ given by the triangle diagram of Fig.~\ref{fig-2} (a). The vertical dashed lines indicate the masses of $P_c^+(4380)$ and $P_c^+(4450)$, respectively. Here the contribution of $\chi_{c2}$ is estimated as the order $10^{-1}$ of that from $\chi_{c1}$.}
\label{fig-chic}
\end{figure}

In Fig.~\ref{fig-chic} we show the structures in the invariant mass of $J/\psi p$ via the triangle diagram of Fig.~\ref{fig-2} (a). As discussed before, since $\chi_{c1} p$ and $\chi_{c2} p$ can access the possible quantum numbers via the $S$ wave, we only consider loops of $\chi_{c1}$ and $\chi_{c2}$ at this moment. On the other hand, since the branching ratio of $B^+\to K^+\chi_{c2}$ is one order of magnitude smaller than that of $B^+\to K^+\chi_{c1}$, we expect that it is also the case in the $\Lambda_b$ decay.  Then, the upper limit of the contributions from $\chi_{c2}$ can be estimated by requiring $BR(\Lambda_b\to K^-\chi_{c2} p)/BR(\Lambda_b\to K^-\chi_{c1} p)\sim 10^{-1}$. As shown in Fig.~\ref{fig-chic}, the ATS can produce significant threshold enhancement of $\chi_{c1}p$ while the effect from the $\chi_{c2} p$ loop is strongly suppressed. Note that the singularity from the $\chi_{c1}p$ loop exactly locates at the mass position of the observed $P_c(4450)$.
The pole trajectories of the $\chi_{c1}p$ loop and the properties of the poles on different Riemann Sheet were first considered in Ref.~\cite{Guo:2015umn}.

\begin{table}
\caption{The $\chi_{cJ}p$ thresholds which can be enhanced by the ATS via Fig.~\ref{fig-2} (a). }\label{tab-chic}
\begin{center}
\begin{tabular}{|c|c|c|c|}
	\hline
Threshold masses [GeV] & $\chi_{c0}(1P) \ 0^+$ & $\chi_{c1}(1P) \ 1^+$ & $\chi_{c2}(1P) \ 2^+$  \\	
\hline
$p \ 1/2^+$ & 4.353  & 4.449  & 4.494   \\
\hline
\end{tabular}
\end{center}
\end{table}

\begin{figure}[t]
\centering
\includegraphics[width=0.95\textwidth]{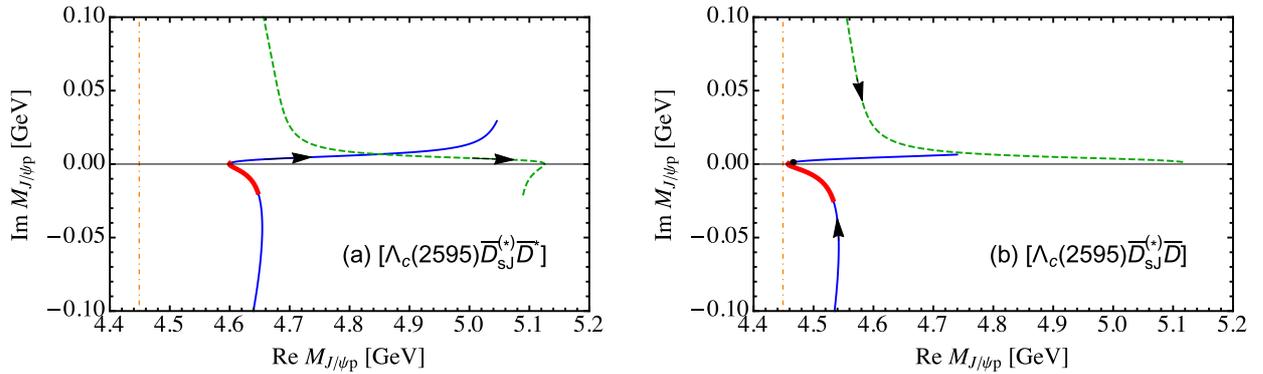}
\caption{ The trajectories of the two singularities of Fig.\ref{fig-2} (b) in terms of the mass of $\bar{D}_{sJ}^{(*)}$ with $m_{\Lambda_c}=2.595~\mathrm{GeV}$. The emitted charmed mesons from $\bar{D}_{sJ}^{(*)}$ are $\bar{D}^*$ (a) and $\bar{D}$ (b), respectively.  The masses of $\bar{D}_{sJ}^{(*)}$ decrease along the arrows. In order to distinguish the pinch singularities, a small width, $10~\mathrm{MeV}$, is assigned to $\bar{D}_{sJ}^{(*)}$.  The black point in (b) corresponds to the $D_{sJ}(2860)$. The red thick solid lines are the possible mass regions which can produce the ATS enhancement.  The vertical dot-dashed lines in both plots indicate the mass position of the $P_c^+ (4450)$. The corresponding trajectory of Fig.~\ref{fig-2} (a) has been investigated in Ref.~\cite{Guo:2015umn}. }
\label{fig-6}
\end{figure}

It is possible that the intermediate $\bar c s$ in Fig.~\ref{fig-1} (b) can form intermediate ${\bar D}_{sJ}^{(*)}$ states which can decay into $\bar D^{(*)}$ and $K^-$. The intermediate $\bar D^{(*)}$ meson will then scatter the $\Lambda_c$ into $J/\psi p$. This process is illustrated in Fig.~\ref{fig-2} (b). The accessible thresholds are listed in Table~\ref{tab-1}. Although one can see from Table~\ref{tab-1} that in an $S$ wave none of the thresholds matches the experimental measured masses and favored quantum numbers simultaneously.  However, among these `$\Lambda_c$'s, the $\Lambda_c(2595)$+$\bar{D}_{sJ}(2860)$ loop with $\bar{D}_{sJ}(2860)$ decay to $\bar{D}K$ can produce the singularity at the $P^+_c(4450)$ mass position as shown by the black point in Fig.~\ref{fig-6} (b).  As a comparison we also show the pole trajectory with $\bar{D}_{sJ}(2860)$ decay to $\bar{D}^*K$ in Fig.~\ref{fig-6} (a).  Note that the ATS only works in a very limited kinematic region. It implies that if the kinematics deviate from the ATS condition one should not expect any significant enhancement at the corresponding threshold mass region. This will  provide a possibility for experimentalists to pin down the nature of some threshold states. Namely, if they are not caused by kinematic effects the enhancements will still appear in other processes where the ATS condition does not hold. 

Although $\Lambda_c(2595)\bar{D}_{sJ}(2860)$ can access the expected quantum number in a $P$ wave which will be suppressed by the centrifugal barrier,
we will show later that the singular points near threshold can still match the observed peak positions.
Also, as mentioned earlier that the ATS can still possibly produce observable effects when higher partial waves are present at the interaction vertices, we then investigate the possible partial waves for Fig.~\ref{fig-2} (b) and see how the ATS would manifest in the invariant mass spectrum.

In Table~\ref{tab-2} the thresholds for the $\Lambda_c^{(*)}{\bar D}_{sJ}^{(*)} $ around the mass of $\Lambda_b$ are listed. To fully satisfy the ATS condition, the threshold of $ \Lambda_c^{(*)}{\bar D}_{sJ}^{(*)}$ should be close to the mass of $\Lambda_b$. We note that for most of the ${\bar D}^{(*)}_{sJ} \Lambda_c^{(*)}$ thresholds, they have already allow a sizeable enhancement near the $\Lambda_c^{(*)} \bar D^{(*)}$ thresholds.

\begin{figure}[tb]
	\centering
	\includegraphics[width=0.4\hsize]{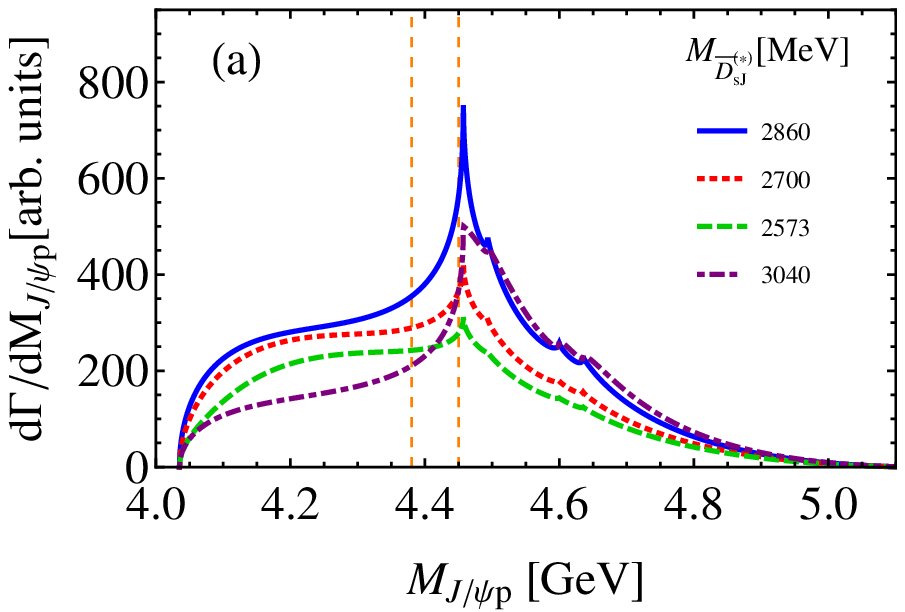}\\
	\includegraphics[width=0.4\hsize]{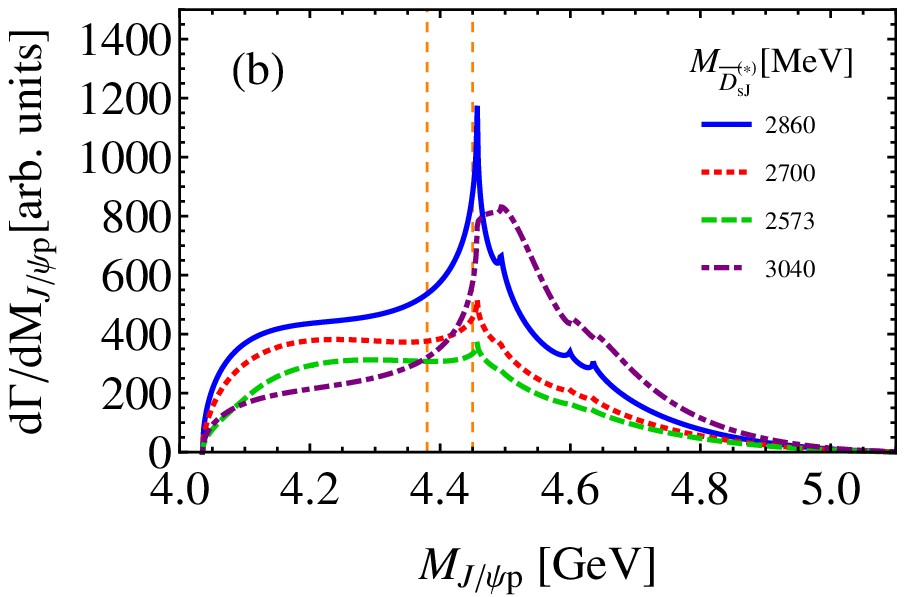}\\
	\includegraphics[width=0.4\hsize]{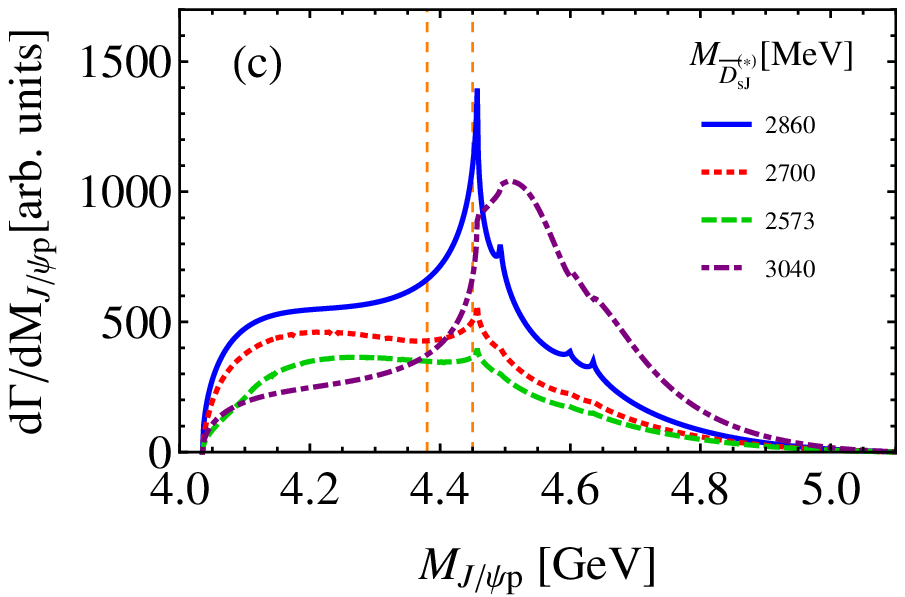}
\caption{Invariant mass distributions of  $J/\psi p$ given by the triangle diagram of Fig.~\ref{fig-2} (b). The width of the intermediate $\bar{D}_{sJ}^{(*)}$  are taken as (a) $150$ MeV, (b) $50$ MeV and (c) $0$ MeV, respectively, as a demonstration of the width effects. The masses of the $\bar{D}_{sJ}^{(*)}$ are $2860$ MeV, $2700$ MeV, $2573$ MeV and $3040$ MeV corresponding to $\bar{D}_{sJ}^{(*)}$ states listed in Table~\ref{tab-2}. The vertical dashed lines indicate the masses of $P_c^+(4380)$ and $P_c^+(4450)$, respectively. }\label{fig-lambdac}
\end{figure}

In Fig.~\ref{fig-2} (b), the charmed baryon and anticharmed meson  are chosen as $\Lambda_{c}(2595)$/$\Lambda_{c}(2650)$ and $\bar{D}$/$\bar{D}^*$, respectively, which means that we take into account the contributions from four diagrams. Furthermore, to match the qualitative feature of the experimental observations, we estimate the rescattering amplitudes by assuming relatively smaller branching fractions of $\Lambda_b\to \bar{D}_{sJ}^{(*)} \Lambda_c(2625)$ and $ \bar{D}_{sJ}^{(*)}\to \bar{D}^* K$, compared with those of $\Lambda_b\to \bar{D}_{sJ}^{(*)} \Lambda_c(2595)$ and $ \bar{D}_{sJ}^{(*)}\to \bar{D} K$, respectively.  The numerical results of $J/\psi p$ invariant mass distributions given by the diagrams of Fig.~\ref{fig-2} (b) are displayed in Fig.~\ref{fig-lambdac}. There will be many options for $\bar{D}_{sJ}^{(*)}$, we therefore display the results for several masses and widths of $\bar{D}_{sJ}^{(*)}$.
 Since the intermediate state $\bar{D}_{sJ}^{(*)}$ in Fig.~\ref{fig-2}(b) could be broad, to account for the width effect we adopt a Breit-Wigner-type propagator $[q^2-m_{\bar{D}_{sJ}^{(*)}}^2+im_{\bar{D}_{sJ}^{(*)}}\Gamma_{\bar{D}_{sJ}^{(*)}}]^{-1}$ in the calculation of the loop integrals. The complex mass of an intermediate state will move the ATS from the physical boundary by a small distance. If the width is not extremely large, the ATS will lie close to the physical boundary, and the scattering amplitude can still feel the influence of the singularity.
 In Fig.~\ref{fig-lambdac}, when the mass of $\bar{D}_{sJ}^{(*)}$ is taken around $2.86$ GeV, one notices that four peaks arise in the $J/\psi p$ distributions, which stay in the vicinities of $\Lambda_{c}(2595)\bar{D}$, $\Lambda_{c}(2625)\bar{D}$, $\Lambda_{c}(2595)\bar{D}^*$ and $\Lambda_{c}(2625)\bar{D}^*$ threshold, respectively. The narrow peak around the $\Lambda_{c}(2595)\bar{D}$ threshold can match the structure of $P_c(4450)$ as observed in experiment. When the widths of $\bar{D}_{sJ}^{(*)}$ vary from $0$ up to $150$ MeV, the main feature of the threshold enhancements still hold though the rates will decrease.

\begin{table}
\caption{Thresholds accessible in the invariant mass spectrum of $J/\psi p$. The two numbers in the square bracket have beyond the allowed phase space for $J/\psi p$.}\label{tab-1}
\begin{center}
\begin{tabular}{|c|c|c|c|c|}
	\hline
Threshold masses [GeV] & $\Lambda_c(2286) \ 1/2^+$ & $\Lambda_c(2595) \ 1/2^-$ & $\Lambda_c(2625) \ 3/2^-$ & $\Lambda_c(2880) \ 5/2^+$ \\	
\hline
${\bar D}(1865) \ 0^-$ & 4.151  & 4.457  & 4.493 & 4.746  \\
	\hline
${\bar D}^*(2007) \ 1^-$ & 4.293 & 4.599 & 4.635 & 4.888 \\
\hline
${\bar D}_1(2420) \ 1^+$ & 4.706 & 5.015 & 5.045 & [5.300] \\
\hline
${\bar D}_2(2460) \ 2^+$ & 4.746 & 5.055 & 5.085 & [5.340] \\
\hline
\end{tabular}
\end{center}
\end{table}

\begin{table}
\caption{Thresholds for the ${\bar D}_{sJ}^{(*)} \Lambda_c^{(*)}$. Threshold values in the square bracket are above the mass of $\Lambda_b(5619)$. }\label{tab-2}
\begin{center}
		\begin{tabular}{|c|c|c|c|c|}
			\hline
			Threshold masses [GeV] & $\Lambda_c(2286) \ 1/2^+$ & $\Lambda_c(2595) \ 1/2^-$ & $\Lambda_c(2625) \ 3/2^-$ & $\Lambda_c(2880) \ 5/2^+$ \\	
			\hline
			${\bar D}_s(1968) \ 0^-$ &  4.254  &  4.563  & 4.593 &  4.848  \\
			\hline
			${\bar D}_s^*(2112) \ 1^-$ & 4.398  & 4.707  &  4.737 & 4.994  \\
			\hline
			${\bar D}_{s0}(2317) \ 0^+$ & 4.585 & 4.912 & 4.942 & 5.197 \\
			\hline
			${\bar D}_{s1}(2460) \ 1^+$ & 4.728 & 5.055 & 5.085 & 5.340 \\
			\hline
			${\bar D}_{s1}(2536) \ 1^+$ & 4.822 & 5.131 & 5.161 & 5.416 \\
			\hline
			${\bar D}_{s2}(2573) \ ?^?$ & 4.859 & 5.168 & 5.198 & 5.453  \\
			\hline
			${\bar D}_{s1}(2700) \ 1^-$ & 4.986 & 5.295 & 5.325 & 5.580 \\
			\hline
			${\bar D}_{sJ}(2860) \ ?^?$ & 5.146 & 5.455 & 5.485 & [5.740] \\
			\hline
			${\bar D}_{sJ}(3040) \ ?^?$ & 5.331 & [5.636] & [5.672] & [5.926] \\
			\hline			
		\end{tabular}
\end{center}
\end{table}

\begin{table}
\caption{The $\Sigma_c^{(*)} \bar D^{(*)}$ thresholds accessible in the invariant mass spectrum of $J/\psi p$. The two numbers in the square bracket have beyond the allowed phase space for $J/\psi p$.}\label{tab-3}
\begin{center}
\begin{tabular}{|c|c|c|c|}
	\hline
Threshold masses [GeV] & $\Sigma_c(2455) \ 1/2^+$ & $\Sigma_c(2520) \ 3/2^+$ & $\Sigma_c(2800) \ ?^?$  \\	
\hline
${\bar D}(1865) \ 0^-$ & 4.321  & 4.385  & 4.668   \\
	\hline
${\bar D}^*(2007) \ 1^-$ & 4.463 & 4.527 & 4.810 \\
\hline
${\bar D}_1(2420) \ 1^+$ & 4.875 & 4.939 & [5.222]  \\
\hline
${\bar D}_2(2460) \ 2^+$ & 4.917 & 4.981 & [5.264]  \\
\hline
\end{tabular}
\end{center}
\end{table}

For Fig.~\ref{fig-2} (c), the kinematic effects will be just CUSP structures in the $J/\psi p$ invariant mass spectrum. We do not discuss the dynamic consequences if the intermediate $\Sigma_c^{(*)}\bar D^{(*)}$ (Table~\ref{tab-3}) may have strong couplings, then they may generate dynamic poles near threshold after proper summation over the bubble loops. Instead, we only show the kinematic CUSP for which has turned out not to lead to pronounced structures as recently studied in Ref.~\cite{Guo:2014iya}. In another word, the observed pronounced peaks may either be produced by possible pole structures or the ATS mechanisms.

\begin{figure}[t]
\centering
\includegraphics[width=0.6\textwidth]{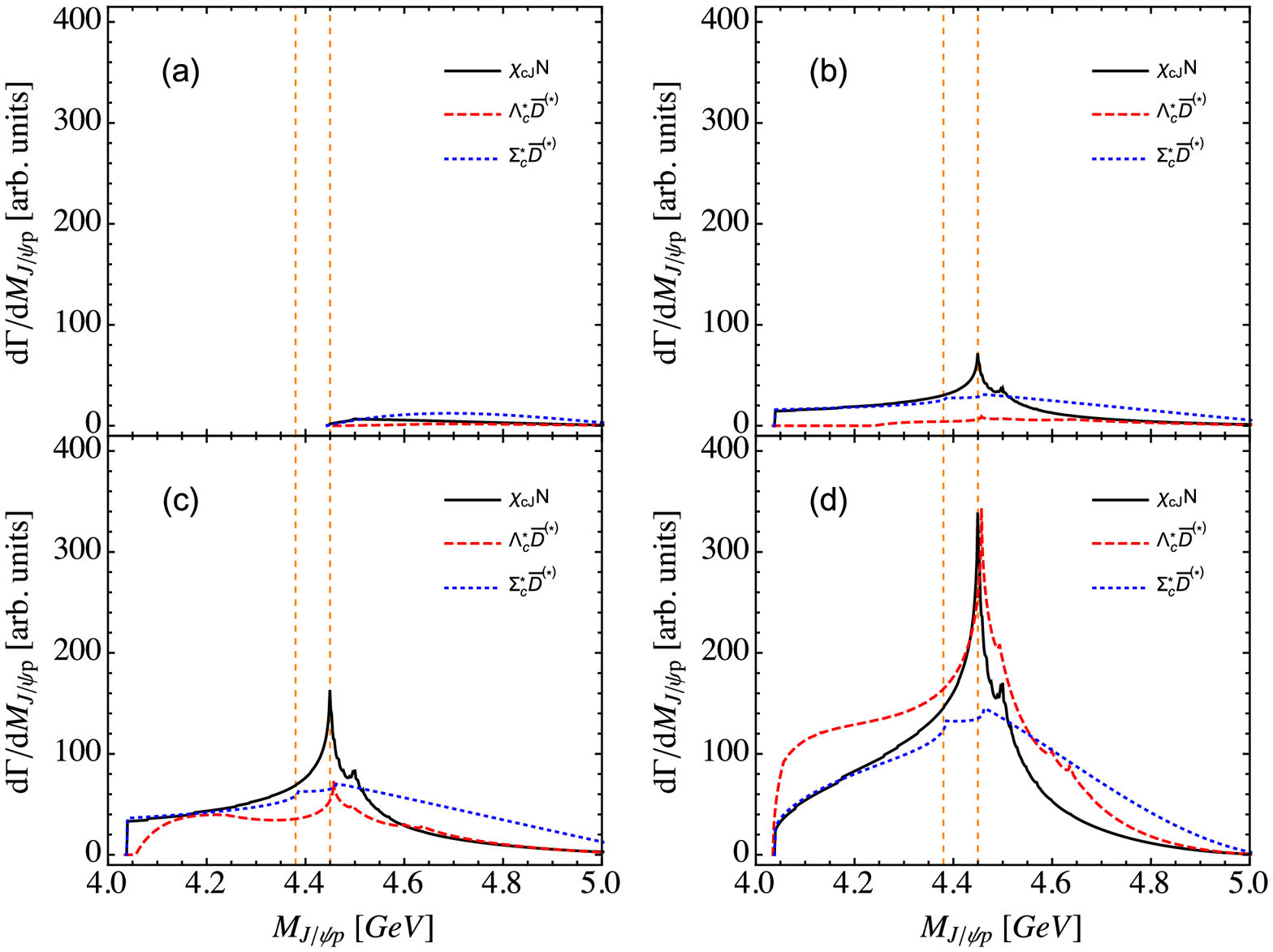}
\caption{Invariant mass distribution of $J/\psi p$ with different $K^- p$ momentum cuts which are same as Fig. 8 of Ref.~\cite{Aaij:2015tga}, i.e. (a) $m_{Kp}< 1.55 \ \mathrm{GeV}$,(b) $1.55 \ \mathrm{GeV} <m_{Kp}< 1.07 \ \mathrm{GeV}$, (c) $1.07 \ \mathrm{GeV}< m_{Kp}< 12.0 \ \mathrm{GeV}$, (d) $ m_{Kp}> 2.0 \ \mathrm{GeV}$. The vertical dashed lines indicate the masses of $P_c^+(4380)$ and $P_c^+(4450)$, respectively.}
\label{fig-cutoff}
\end{figure}

In order to try to distinguish the behavior from a pole structure and the ATS, we generate the invariant mass spectra of $J/\psi p$ in different $K^- p$ invariant mass regions the same as Fig. 8 of Ref.~\cite{Aaij:2015tga} (see Fig.~\ref{fig-cutoff}). Since the ATS contributions will vary in terms of different kinematics, such a quantity will be able to distinguish the behavior of the pole and the ATS mechanism. The results for those three loop processes of Fig.~\ref{fig-2} (a), (b) and (c) are illustrated by the solid, dashed and dotted lines which are different from the symmetric Breit-Wigner lineshape. In particular, the structures created by the CUSP effects appear to be negligible. Note that the kinematics of higher invariant mass of $K^- p$ is favored by Fig.~\ref{fig-2} (a). It means even small couplings for $\chi_{cJ} p$ scattering are introduced, the ATS enhancement can still be observable.

Since we introduce several mechanisms to generate the kinematic singularities, it is also necessary to discuss their similar and different characteristics. Both $\chi_{c1}p$ and $\Lambda_c(2595)\bar{D}$ thresholds are very close to the mass of $P_c^+(4450)$.  Note that $\chi_{c1}p$ can scatter into $J/\psi p$ via multi-gluon exchange process, while $\Lambda_c^*\bar{D}$ can scatter into $J/\psi p$ via quark interchange process which is a rearrangement of the quark flavors.   The $P$-wave scattering between $\chi_{c1}$ and proton will imply that the quantum numbers of the $J/\psi p$ system can be $J^P=(1/2^-, \ 3/2^-, \ 5/2^-)$, but the $P$-wave scattering between $\Lambda_c(2595)$ and $\bar{D}$ will imply that the quantum numbers can be $J^P =(1/2^-, 3/2^-)$. Some of these quantum numbers are compatible with the experimental fitting results.  If one hopes the rescattering mechanism can favor the $J^P=5/2^+$ assignment of $P_c^+(4450)$, the $D$-wave rescattering of $\chi_{c1}p$ or $\Lambda_c(2595)\bar{D}$ will be required. Usually the higher partial wave rescattering amplitudes will be suppressed to some extent. However, the analytic properties of the kinematic singularities will mainly depend on the kinematics of the loop integrals, which will not be affected too much by the coupling forms of the vertices. To  clarify whether the $\chi_{cJ}p$ rescatterings or the $\Lambda_{c}^*\bar{D}^{(*)}$ rescatterings would be dominant in producing the resonance-like peaks, we suggest that $\Lambda_b^0\to K^- h_c p$ would be a promising channel. Namely, the transition $\chi_{cJ}p \to h_c p$ would require the heavy quark spin flip, thus, will be suppressed. In contrast, in  $\Lambda_{c}^*\bar{D}^{(*)}$$\to$$h_c p$ such a spin flip does not necessarily occur. Furthermore, since the $h_c p$ threshold is much larger than the lower thresholds, some higher resonance-like peaks induced by the rescatterings, similar to $P_c^+(4450)$, can be expected in the $h_c p$ invariant mass distributions.

\section{Summary}

In this work we analyze the role played by anomalous triangle singularity in $\Lambda_b\to J/\psi K^- p$ as a possible contribution to the observed hidden-charm pentaquark candidates $P_c^+(4380)$ and $P_c^+(4450)$. We first show that the $\Sigma_c^{(*)} \bar D^{(*)}$ can be produced at leading order which seems to be overlooked in the literature. We then demonstrate that the ATS can generate threshold enhancements which can mimic the experimental observations. We have to admit that more detailed model construction is needed in order to determine better the ATS behavior and strengths. But we emphasize that the kinematics of $\Lambda_b\to J/\psi K^- p$ are in favor of the ATS mechanism when intermediate states are involved. Since all the triangle amplitudes can build up the contributions from the ATS near the mass region that we are interested in, they seem to be able to produce structures similar to what observed in experiment. Meanwhile, we find that the $\Lambda_c(2625) \bar D(1865)$ threshold can give rise to a third peak around 4.5 GeV.

We also discuss the kinematic feature of $\Sigma_c^{(*)}\bar D^{(*)}$ which can produce kinematic CUSP effects. In this sense, the observed pronounced structures can be either produced by pole structures or the ATS mechanism. The kinematic dependence of the ATS is investigated by looking at the ATS cross sections in different $K^- p$ invariant mass regions. We also show that the ATS mechanism should have rather strong dependence of the kinematics by looking at the $J/\psi p$  invariant mass distributions at different energy cuts for $K^-p$.

In brief, since the ATS could play a role in the production of threshold states,   it may mix with the pole if the singular threshold is close to the pole position. Such a mixing and interference might generate complicated structures such as to distort the lineshape or produce narrower enhancements that behaves very differently from expected pole contributions.  In order to better understand the nature of these newly observed pentaquark candidates, a combined study of the ATS and dynamically generated pole structures should be necessary.

\section*{Acknowledgement}

Useful discussions with Ulf-G. Mei{\ss}ner and  M. Oka are acknowledged. This work is supported, in part, by the Japan Society for the Promotion of Science under Contract No. P14324, the JSPS KAKENHI (Grant No. 25247036), the Sino-German CRC 110 ``Symmetries and
the Emergence of Structure in QCD" (NSFC Grant No. 11261130311), the National Natural Science Foundation of China (Grant Nos. 11425525), and the National Key Basic Research Program of China under Contract No. 2015CB856700.


\begin{thebibliography}{99}


\bibitem{Aaij:2015tga}
  R.~Aaij {\it et al.} [LHCb Collaboration],
  arXiv:1507.03414 [hep-ex].  

\bibitem{Wu:2010jy}
  J.~J.~Wu, R.~Molina, E.~Oset and B.~S.~Zou,
  Phys.\ Rev.\ Lett.\  {\bf 105}, 232001 (2010)  [arXiv:1007.0573 [nucl-th]].


\bibitem{Wu:2010vk}
  J.~J.~Wu, R.~Molina, E.~Oset and B.~S.~Zou,
  Phys.\ Rev.\ C {\bf 84}, 015202 (2011)  [arXiv:1011.2399 [nucl-th]].


\bibitem{Wu:2012md}
  J.~J.~Wu, T.-S.~H.~Lee and B.~S.~Zou,
  Phys.\ Rev.\ C {\bf 85}, 044002 (2012)  [arXiv:1202.1036 [nucl-th]].

\bibitem{Yang:2011wz}
  Z.~C.~Yang, Z.~F.~Sun, J.~He, X.~Liu and S.~L.~Zhu,
  Chin.\ Phys.\ C {\bf 36}, 6 (2012)  [arXiv:1105.2901 [hep-ph]].

\bibitem{xiao-2013}
C.~W.~Xiao, J.~Nieves and E.~Oset,
Phys.\ Rev.\ D {\bf 88}, 056012 (2013).


\bibitem{Choi:2007wga}
  S.~K.~Choi {\it et al.} [Belle Collaboration],
  Phys.\ Rev.\ Lett.\  {\bf 100}, 142001 (2008)  [arXiv:0708.1790 [hep-ex]].


\bibitem{Ablikim:2013mio}
M.~Ablikim {\it et al.}  [BESIII Collaboration],
Phys.\ Rev.\ Lett.\  {\bf 110}, 252001 (2013)
[arXiv:1303.5949 [hep-ex]].


\bibitem{Guo:2014iya}
F.~K.~Guo, C.~Hanhart, Q.~Wang and Q.~Zhao,
Phys.\ Rev.\ D {\bf 91}, no. 5, 051504 (2015)
[arXiv:1411.5584 [hep-ph]].



\bibitem{Chen:2015loa}
  R.~Chen, X.~Liu, X.~Q.~Li and S.~L.~Zhu,
  arXiv:1507.03704 [hep-ph].


\bibitem{Chen:2015moa}
  H.~X.~Chen, W.~Chen, X.~Liu, T.~G.~Steele and S.~L.~Zhu,
  arXiv:1507.03717 [hep-ph].



\bibitem{Karliner:2015ina}
  M.~Karliner and J.~L.~Rosner,
  arXiv:1506.06386 [hep-ph].

\bibitem{Roca:2015dva}
  L.~Roca, J.~Nieves and E.~Oset,
  arXiv:1507.04249 [hep-ph].

\bibitem{Feijoo:2015cca}
  A.~Feijoo, V.~K.~Magas, A.~Ramos and E.~Oset,
  arXiv:1507.04640 [hep-ph].



\bibitem{Liu:2015taa}
  X.~H.~Liu, M.~Oka and Q.~Zhao,
  arXiv:1507.01674 [hep-ph].  




\bibitem{Wu:2011yx}
J.~J.~Wu, X.~H.~Liu, Q.~Zhao and B.~S.~Zou,
Phys.\ Rev.\ Lett.\  {\bf 108}, 081803 (2012)
[arXiv:1108.3772 [hep-ph]].


\bibitem{Wu:2012pg}
X.~G.~Wu, J.~J.~Wu, Q.~Zhao and B.~S.~Zou,
Phys.\ Rev.\ D {\bf 87}, no. 1, 014023 (2013)
[arXiv:1211.2148 [hep-ph]].



\bibitem{Wang:2013cya}
Q.~Wang, C.~Hanhart and Q.~Zhao,
Phys.\ Rev.\ Lett.\  {\bf 111}, no. 13, 132003 (2013)
[arXiv:1303.6355 [hep-ph]].


\bibitem{Wang:2013hga}
Q.~Wang, C.~Hanhart and Q.~Zhao,
Phys.\ Lett.\ B {\bf 725}, no. 1-3, 106 (2013)
[arXiv:1305.1997 [hep-ph]].



\bibitem{Liu:2013vfa}
X.~H.~Liu and G.~Li,
Phys.\ Rev.\ D {\bf 88}, 014013 (2013)
[arXiv:1306.1384 [hep-ph]].




\bibitem{Liu:2014spa}
X.~H.~Liu,
Phys.\ Rev.\ D {\bf 90}, no. 7, 074004 (2014)
[arXiv:1403.2818 [hep-ph]].

\bibitem{Guo:2015umn}
  F.~K.~Guo, Ulf~-G.~Mei{\ss}ner, W.~Wang and Z.~Yang,
  arXiv:1507.04950 [hep-ph].

\bibitem{Landau:1959fi}
L.~D.~Landau,
Nucl.\ Phys.\  {\bf 13}, 181 (1959).


\bibitem{bonnevay:1961aa}
G.~Bonnevay, I.~J.~R.~Aitchison and J.~S.~Dowker,
Nuovo\ Cim.\  {\bf 21}, 1001 (1961).



\end{thebibliography}
\end{document}